\documentclass[12pt]{article}
\usepackage{pic03}
\usepackage{hyperref}
\usepackage{url}
\usepackage{graphicx}

\begin{document}

\title{\bf Heavy flavour production with leptons and hadrons}
\author{
Matthew Wing        \\
{\em Bristol University, ZEUS, DESY, Notkestrasse 85, 22607 Hamburg, Germany}}
\maketitle

%
% photograph of author
%  This is where we will insert a photograph. To see what it would look like,
%  uncomment the following lines.
%
%\begin{figure}[h]
%\begin{center}
%
% include photograph for proceeding version
%
%\includegraphics
%[height=4.5cm]{include_figures/wing.eps}
%
% insert a fixed vertical spacing instead for the ArXiv preprint
%
\vspace{4.5cm}
%
%\end{center}
%\end{figure}

\baselineskip=14.5pt
\begin{abstract}
The production of charm and beauty quarks in $\gamma \gamma$ collisions at LEP, 
$ep$ collisions at HERA and $p \bar{p}$ collisions at the Tevatron is discussed. 
The comparison with predictions of next-to-leading-order QCD 
and the issues it raises are detailed. In particular, the strengths and weaknesses 
of the measurements and predictions are discussed.
\end{abstract}
\newpage

\baselineskip=17pt

\section{Introduction}

Measurements of the production of heavy quarks provide a wealth of information 
on high energy particle collisions. The mechanism for production of heavy 
quarks is governed by the strong force of nature which is described by Quantum 
Chromodynamics (QCD). Not only is QCD essential for understanding one of the 
four fundamental forces of nature, but many signatures of physics beyond the 
Standard Model (SM) are dependent on precise knowledge of the rate of QCD 
processes, which are expected to form the most significant background.

\begin{figure}[htb]
\begin{center}
\includegraphics[height=6.cm]{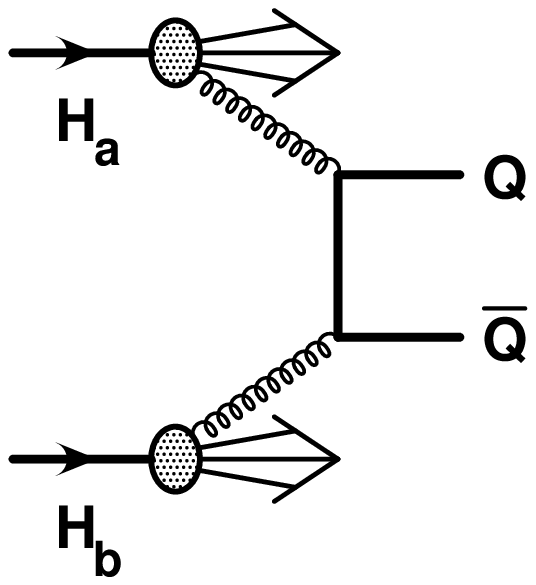}
\hspace{2.5cm}\includegraphics[height=6.cm]{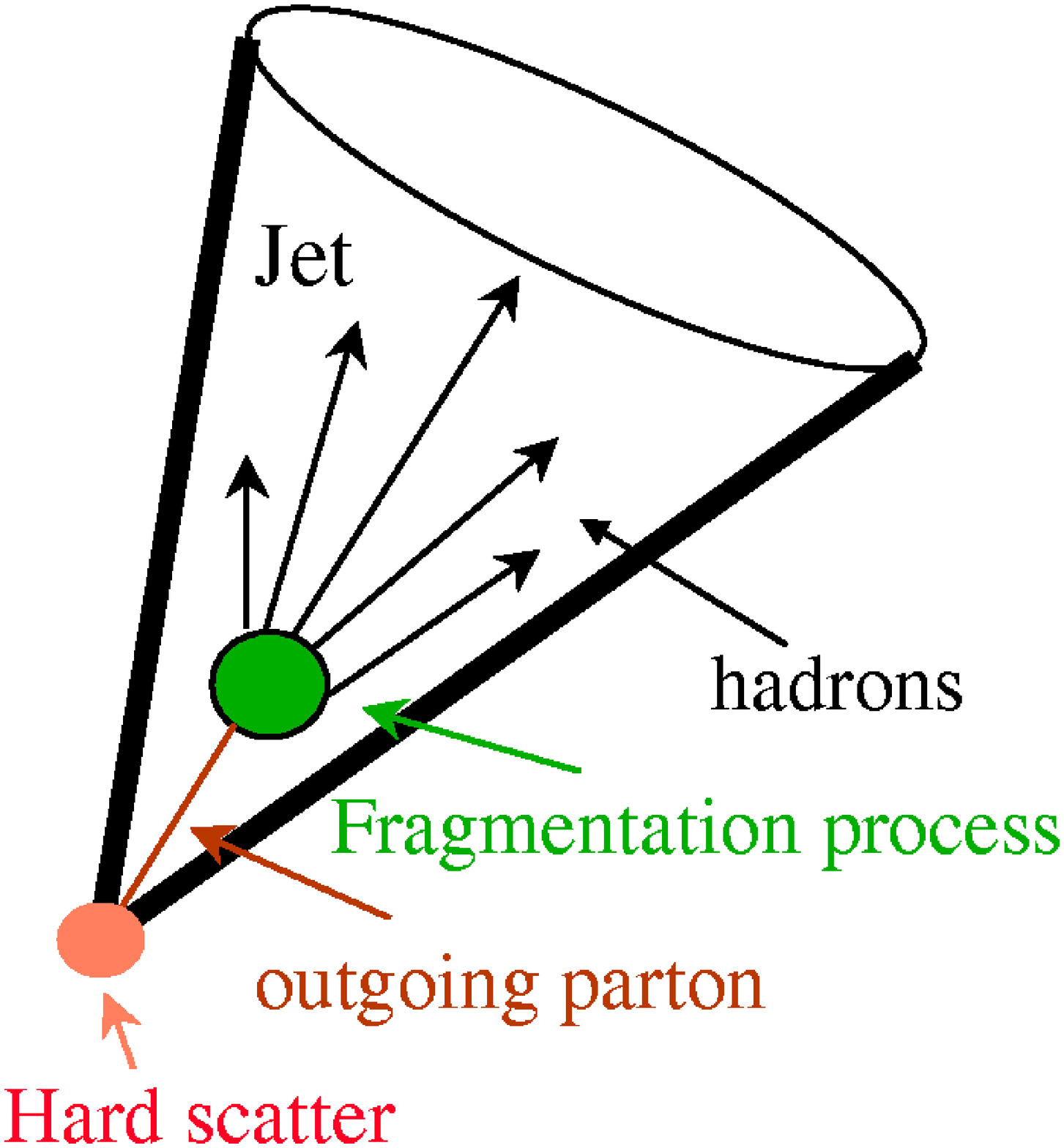}
\put(-220,160){\makebox(0,0)[tl]{\bf (a)}}
\put(-30,160){\makebox(0,0)[tl]{\bf (b)}}
\end{center}
\caption{\it Generic representation of (a) the production of heavy quarks 
in hadron-hadron collisions and (b) their subsequent hadronisation.
\label{fig:feyn1} }
\end{figure}

The importance of a precise understanding of QCD is apparent when considering 
current and future accelerators. Many of these accelerators will use protons 
and photons as the colliding particles, both of which have a hadronic structure 
and hence are described by QCD. Figure~\ref{fig:feyn1}(a) shows a generic 
representation of the production of heavy quarks in a hadron-hadron collision. 
Knowledge accumulated at HERA, LEP and the Tevatron will directly 
benefit future programmes such as the LHC and a future linear collider where 
heavy quarks will be produced by the same mechanism. The produced partons 
then fragment into final-state hadrons which are measured in the detector as 
depicted in Figure~\ref{fig:feyn1}(b). 
The fragmentation procedure is usually described by non-perturbative models and 
is again an uncertainty common to all experiments. 

Theoretically, heavy quarks provide ideal tools for 
probing QCD due to their relatively large mass, $m_Q >> \Lambda_{\rm QCD}$, 
which entails a fast convergence of the perturbative expansion of the cross 
section. The production of heavy quarks is also {\em directly} sensitive 
to the gluon density in the colliding hadron (see Figure~\ref{fig:feyn1}(a)). 
The gluon density is usually determined in the DGLAP-evolution fits 
to measurements of structure functions in inclusive deep-inelastic scattering. 
Direct measurements of the gluon density provide an important check of 
these methods and the factorisation of the cross section.

In these proceedings, results of open charm and beauty production from HERA, 
the Tevatron and $\gamma \gamma$ collisions at LEP are discussed. After a brief 
overview of theoretical and experimental aspects, emphasis is 
given to understanding hadronic structure and the dynamics of the hard scatter. 
Results on quarkonia and top production and measurements of $B$ fragmentation 
functions are discussed elsewhere~\cite{other_procs}.

\section{Perturbative QCD}

For the generic collision, shown in Figure~\ref{fig:feyn1}, of two hadrons 
producing heavy quarks, $H_a + H_b \to Q \bar{Q} + X$, the cross section can 
be written as a convolution of the parton densities, $f_i^{H_a}$ and 
$f_j^{H_b}$, of the two hadrons and the short-distance cross section, 
$\sigma_{ij}$:
\begin{equation}
\sigma(S) = \sum_{i,j} \int dx_1 \int dx_2 \ \hat{\sigma}_{ij}(x_1 x_2 S, m^2, \mu^2)
                                            f_i^{H_a}(x_1, \mu) f_j^{H_b}(x_2, \mu)
\end{equation}
where $x_1$ and $x_2$ are the hadron's momentum fraction carried by the 
interacting parton and $\mu$ represents the renormalisation and factorisation 
scales. The short-distance cross section is a perturbative expansion in powers 
of $\alpha_s$ and the inverse mass of the heavy quark which implies faster 
convergence for larger masses. Two different schemes and their combination 
are used for predictions of heavy-quark production. In the ``massive'' scheme, 
there are no heavy quarks in the colliding hadron and the predictions should be 
more accurate for transverse momenta $p_T \sim m$. In contrast, heavy quarks in 
the ``massless'' scheme 
are active in the colliding hadrons and the predictions should be more 
accurate for $p_T > m$. The two schemes have recently been 
combined~\cite{fonll,b_cacc_nason} such that predictions should be appropriate 
for all $p_T$.

\section{Experimental techniques}

\begin{figure}[htb]
\begin{center}
\includegraphics[height=5.75cm]{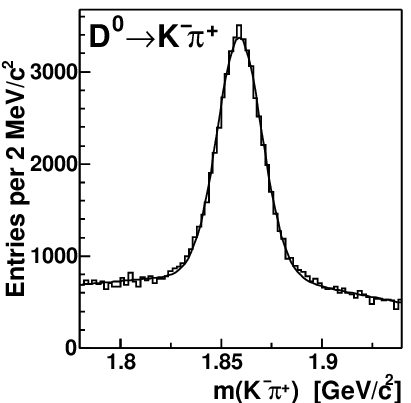}
\includegraphics[height=6.25cm]{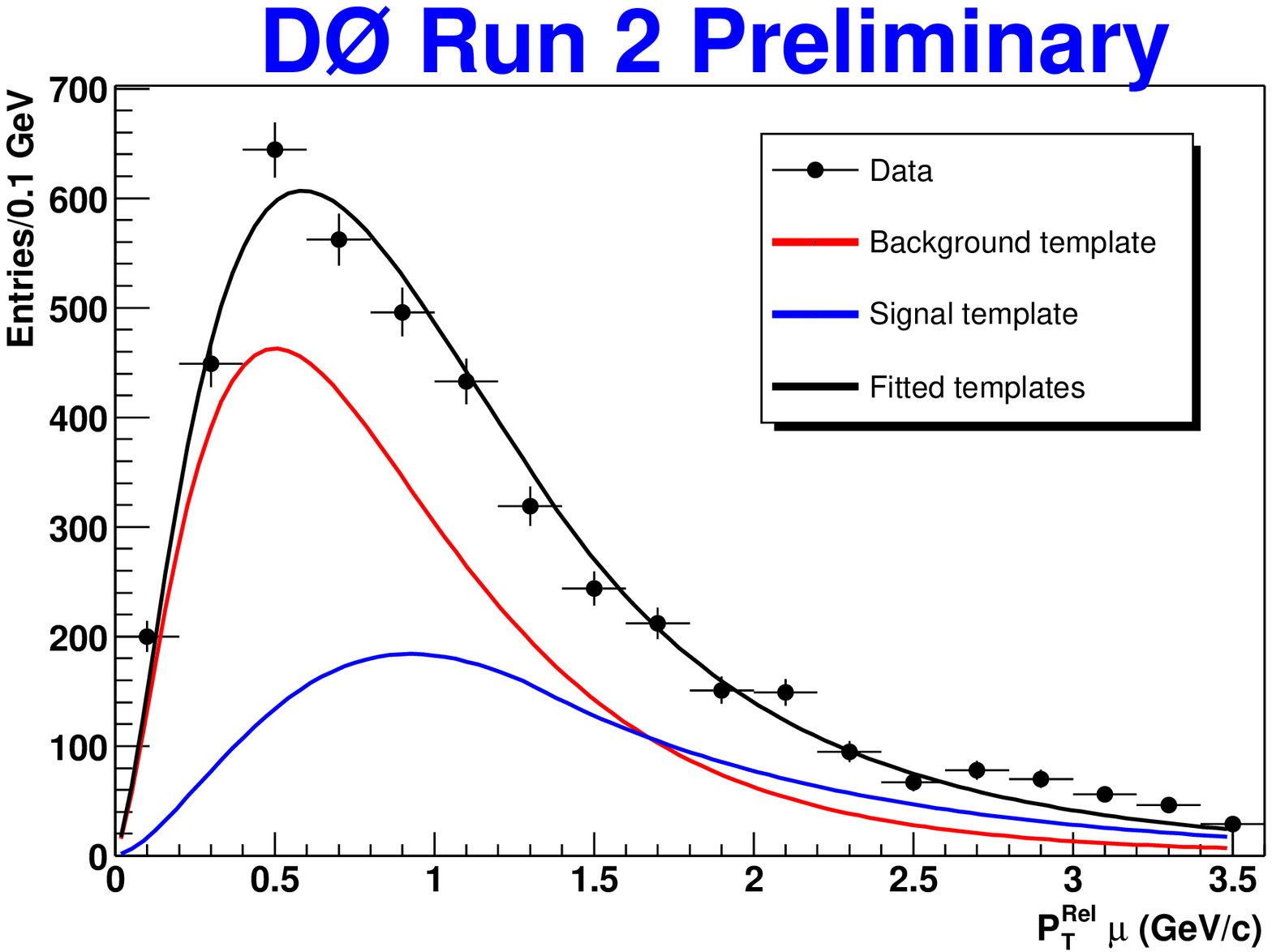}
\put(-280,90){\makebox(0,0)[tl]{\bf (a)}}
\put(-30,90){\makebox(0,0)[tl]{\bf (b)}}
\end{center}
\vspace{-0.5cm}
\caption{\it Methods for tagging (a) charm quarks via the reconstruction of 
a $D^0$ meson and (b) beauty quarks using $p_T^{\rm rel}$.
\label{fig:charms_tev} }
\end{figure}

The reconstruction of heavy quark mesons is similar in all experiments. Signals 
for charm hadrons are generally observed by forming the invariant mass of the 
tracks identified with a specific decay channel, e.g. $D^0 \to K^- \pi^+$. If 
sufficiently accurate, a vertex detector can also be used to detect vertices 
displaced from the primary interaction point. An example of reconstructed $D^0$ 
mesons which uses both these techniques is shown in Figure~\ref{fig:charms_tev}(a) 
from the CDF experiment~\cite{charm-tev}. To detect beauty quarks, 
the invariant mass of the decay products is also formed for a specific decay 
channel such as $B^+ \to J/\psi K^+$. As the decay length of beauty is longer 
than that for charm quarks, a vertex detector is a powerful tool for 
distinguishing  beauty from the lighter quarks. The transverse momentum of an 
electron or muon relative to the direction of the parent quark, $p_T^{\rm rel}$ 
also provides a clear signature. Due to its larger mass, $b$ quarks populate 
high values of $p_T^{\rm rel}$ as shown in 
Figure~\ref{fig:charms_tev}(b)~\cite{d0-moriond}.

Due to limitations of the experimental apparatus, measurements are performed 
in a restricted kinematic region, usually defined by some momentum and angular 
restriction of the reconstructed heavy quark meson. Measurements performed in 
a restricted kinematic region are often extrapolated to the full phase space 
using either a next-to-leading order (NLO) calculation or Monte Carlo (MC) 
model. These extrapolations provide ``measurements'' of total cross sections 
or structure functions which are more intuitive and easy to compare between 
different experiments. They should, however, be treated with 
caution. The extrapolation is often performed to completely unmeasured 
regions, with factors as high as 20 and an uncertainty which is difficult to 
determine.

\section{Latest results}

\begin{figure}[htb]
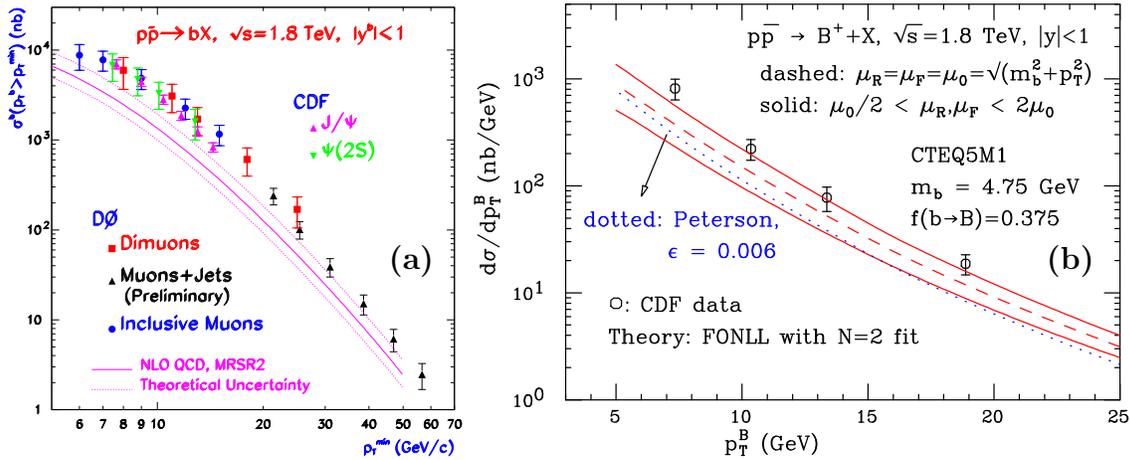

\begin{center}
\includegraphics[height=6.cm]{include_figures/CDF-D0_b_Xsect.epsi}
\includegraphics[height=6.05cm]{include_figures/B-hadrons.epsi}
\put(-280,80){\makebox(0,0)[tl]{\bf (a)}}
\put(-35,80){\makebox(0,0)[tl]{ \bf (b)}}
\end{center}
\vspace{-0.5cm}
\caption{\it Measurements of (a) the $b$ quark cross section compared to NLO 
predictions and (b) the $B^+$ meson cross section compared with FONLL predictions 
from the Tevatron.
\label{fig:b_tev_1} }
\end{figure}

Measurements of the beauty cross section using Tevatron Run I 
data~\cite{b_tev_run1}, shown in 
Figure~\ref{fig:b_tev_1}, provoked much of the current interest in the production 
rate of heavy quarks. Several decay channels were analysed. These results were 
then extrapolated to a cross section for some minimum transverse momentum of the 
$b$ quark, $p_T^{\rm min}$; the measurements are shown in 
Figure~\ref{fig:b_tev_1}(a) compared with an NLO prediction~\cite{pp-nlo}. The 
NLO prediction lies significantly below the data by a factor of 2--3 and is, 
therefore, one of the most significant failures of pQCD to describe high energy 
phenomena of the strong interaction.
In order to remove unknowns associated with the extrapolation procedure, measurements 
of $B$ mesons, which are directly reconstructed in the detector, were performed 
and compared with NLO QCD. As with the cross sections for the $b$ quark, the NLO 
prediction was significantly below the data. With the extra information provided 
by measuring the meson cross section, significant theoretical development was 
made~\cite{b_cacc_nason} to try and describe these data. The improved theoretical 
calculations include the resummation of large logarithms in $p_T$ at the 
next-to-leading level (NLL) and their merging with the fixed order (FO) calculation 
which correctly accounts for mass effects. This new ``FONLL'' calculation also 
uses a new extraction of the fragmentation function of $b$ quarks to $B$ mesons as 
measured in an $e^+e^-$ experiment~\cite{frag-aleph}. The result of this new 
calculation is shown 
compared to the data in Figure~\ref{fig:b_tev_1}(b), where the difference between 
data and theory is reduced from a factor of 2.9 to 1.7 and consistency within the 
uncertainties is observed.

\subsection{Structure functions}

\begin{figure}[htp]
\begin{center}
\includegraphics[width=8.55cm]{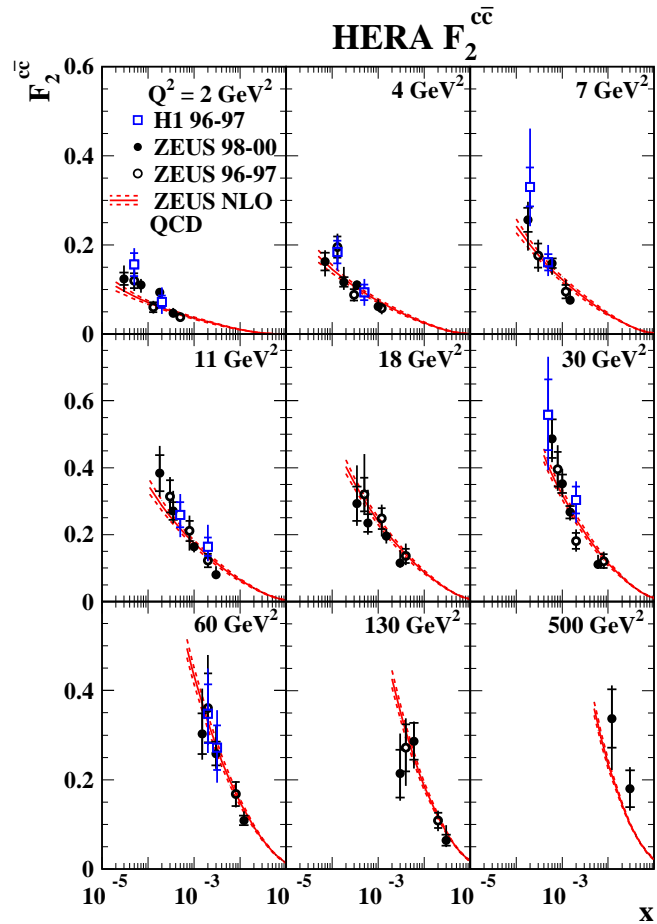}
\end{center}
\vspace{-0.5cm}
\caption{\it Charm contribution, $F_2^{c \bar{c}}$, to the proton structure 
function for different $Q^2$.
\label{fig:f2charm} }
\end{figure}

In deep inelastic scattering (DIS) at HERA, photons act as a pointlike probe of 
the proton and hence provides the unique 
opportunity to study the charm contribution, $F_2^{c \bar{c}}$, to the proton 
structure, $F_2$. Measurements of charm in DIS are directly sensitive to the 
gluon distribution in the proton and are therefore complementary to extractions 
of the gluon distribution in QCD fits. The largest sample of events for charm 
in DIS in both H1 and ZEUS~\cite{h1_z_dstar_dis} is tagged using the ``golden'' 
$D^*$ decay channel. 
Due to experimental limitations, the $D^*$ meson is restricted in to the central 
region of the detector with transverse momentum larger than 1.5~GeV. Cross 
sections are measured differentially in $Q^2$ and $x$ and compared with NLO 
QCD. The NLO QCD is then used to extract $F_2^{c \bar{c}}$:
\begin{equation}
F_{2, \rm meas}^{c \bar{c}} = \frac{\sigma(x,Q^2)_{\rm meas}}{\sigma(x,Q^2)_{\rm theo}} 
                               F_{2, theo}^{c \bar{c}}
\label{eq:f2charm}
\end{equation}
The extrapolation factors to the full $D^*$ phase space vary between 4.7 at low 
$Q^2$ and 1.5 at high $Q^2$. The extracted $F_2^{c \bar{c}}$ values are shown 
compared with an NLO QCD prediction in Figure~\ref{fig:f2charm}. 
The data show a steep rise to low $x$ indicative of a large gluon density in the 
proton. The NLO QCD prediction is derived from fits~\cite{mandy_nlo} to 
measurements of $F_2$ and is largely independent of the data shown here. The data 
are well described by the NLO prediction demonstrating the consistency of the the 
gluon density 
extracted in PDF fits and ``measured'' more directly here. At low $Q^2$, the data, 
specifically the double differential cross sections, $\sigma(x,Q^2)_{\rm meas}$, 
measured within the acceptance of the detector, have reached sufficient precision 
such that the can be used to further constrain the gluon density in the proton. 

\begin{figure}[htb]
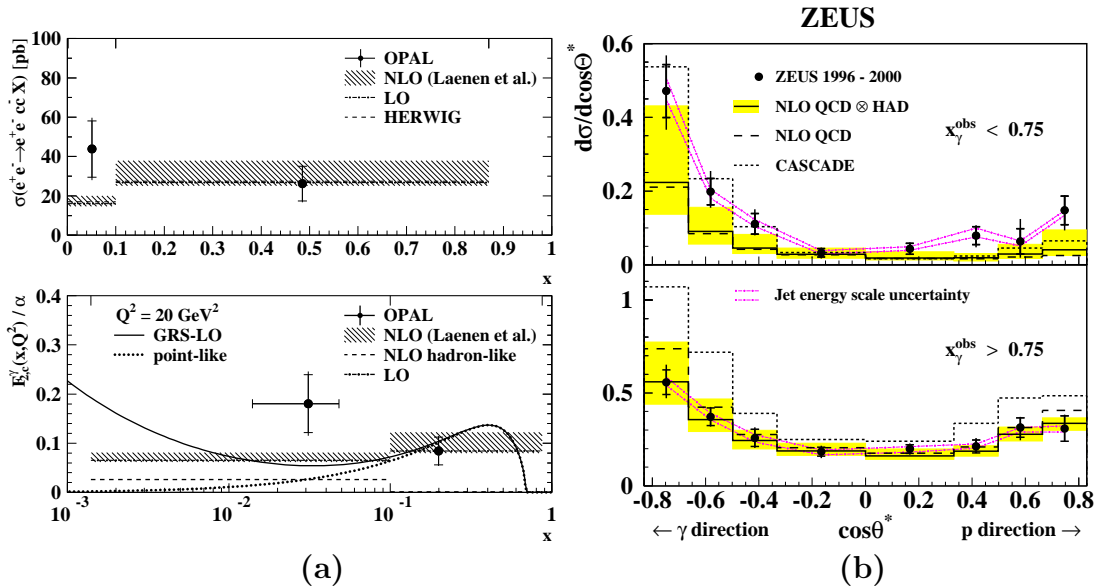

\begin{center}
\includegraphics[width=7.3cm]{include_figures/pr354_04.epsi}
\includegraphics[width=7cm]{include_figures/costheta.epsi}
\put(-300,-5){\makebox(0,0)[tl]{\bf (a)}}
\put(-100,-5){\makebox(0,0)[tl]{ \bf (b)}}
\end{center}
\caption{\it Measurement of (a) charm contribution to the photon structure function 
from LEP and (b) dijet angular distributions in charm photoproduction at HERA.
\label{fig:photon} }
\end{figure}

In an analogous way, measurements of $D^*$ production in $e \gamma$ collisions 
at LEP allow extractions of the charm contribution, $F_{2,c}^\gamma$, to the photon 
structure function. The extraction of $F_{2,c}^\gamma$ is done as in 
Equation~\ref{eq:f2charm} except that the extrapolation of the $D^*$ to the full 
phase space is performed using MC models rather than a NLO calculation. 
Figure~\ref{fig:photon}(a) 
shows the total charm cross section and the extracted $F_{2,c}^\gamma$ compared 
with predictions from NLO QCD. The measurements at high $x$, which are well 
described by NLO, are indicative of the scattering of two pointlike photons. At 
low $x$, the data is somewhat above the prediction in a region where it is expected 
that one of the photons exhibits some hadronic structure~\cite{opal-f2cgamma}.

Jet photoproduction at HERA is a complementary way of studying the structure of the 
photon. A recent measurement~\cite{costheta} of the dijet scattering angle, 
$\theta^*$, in $D^*$ 
production is shown in Figure~\ref{fig:photon}(b). The distribution is sensitive 
to the propagator in the hard scatter and thereby sensitive to the nature of 
the sub-process. The tagged $D^*$ meson is associated with one of the jets and 
the scattering angle of this jet defined with respect to the proton direction. The 
angular distribution, enriched in direct photon processes 
($x_\gamma^{\rm obs} > 0.75$), exhibits a symmetric distribution with a shallow rise 
to high values of $\cos\theta^*$. This is indicative of the exchange of a quark in 
the hard sub-process with the charm produced via the boson-gluon fusion process. At 
low $x_\gamma^{\rm obs}$, where the sample is enriched in resolved photon processes, 
the data are asymmetric, exhibiting a rapid rise to negative $\cos\theta^*$. This 
demonstrates that the charm comes from the photon and exchanges a gluon in the hard 
process. The prediction of NLO in which charm is produced in the hard sub-process and 
is not an active flavour in the structure function, lies below the data. The 
description of the data could be improved by including a charm component in a NLO 
fit of the photon PDF.

\subsection{Measurements of charm cross sections}

Cross sections measured within the acceptance of the detector, such as the angular 
distributions just discussed, do not rely on 
any model assumptions and provides ``safe'' data with which to compare any 
theoretical prediction. Measurements of $D^*$ cross sections are available 
in $\gamma \gamma$ processes at LEP, from the Tevatron and in both DIS and 
photoproduction at HERA. These have been compared with NLO calculations at 
fixed order, or with NLL calculations or the two ``matched''. These produce 
$c \bar{c}$ partons in the final state and incorporate a model of the 
fragmentation into $D^*$ mesons.

Measurements have been performed and compared at LEP by three collaborations, 
ALEPH, L3 and OPAL\cite{alo_dstar}. As a function of $p_T(D^*)$, the data are compatible 
with each other and with FO and NLL calculations. The data is not sufficiently 
precise to distinguish between the different calculations.

\begin{figure}[htb]
\begin{center}
\begin{minipage}[b]{7.5cm}
\includegraphics[height=7.cm]{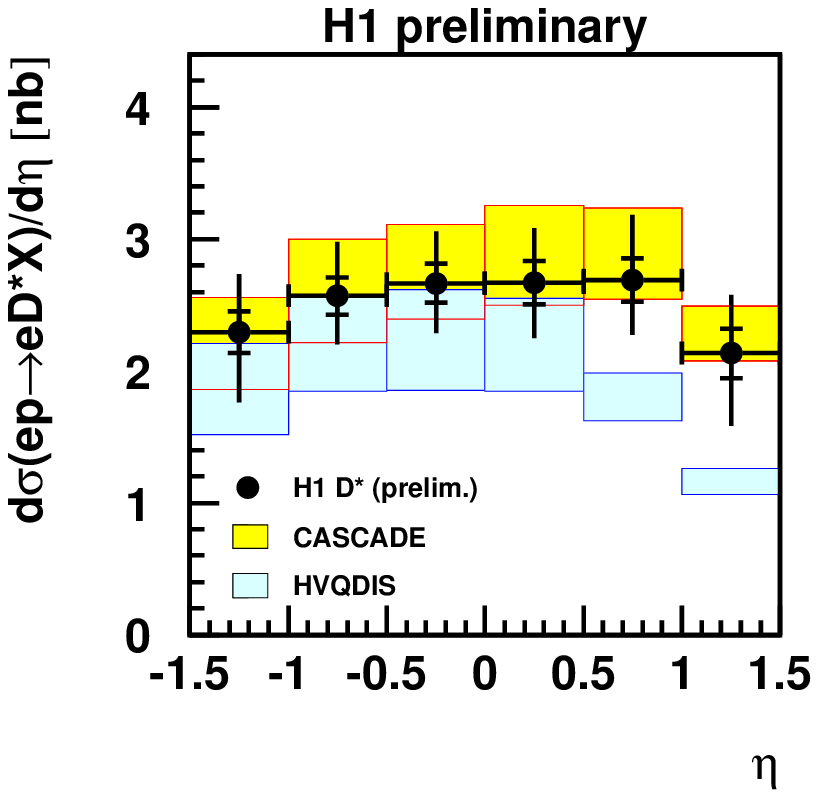}
\vspace{1.5cm}
%\hspace{1.5cm}
\end{minipage}
\begin{minipage}[b]{7.3cm}
\includegraphics[height=9.cm]{include_figures/fig4d_sup.epsi}
\end{minipage}
\end{center}
\vspace{-0.5cm}
\caption{\it Cross sections in DIS from both HERA experiments compared with 
predictions from NLO QCD (and CASCADE MC).
\label{fig:charmx_dis} }
\end{figure}

The HERA data which was used to extract $F_2^{c \bar{c}}$ in the previous 
section are shown in Figure~\ref{fig:charmx_dis}. The data are shown for the same 
variable, $\eta(D^*)$, in a similar kinematic range compared to the calculation, 
{\sc Hvqdis}~\cite{hvqdis}. The NLO calculation lies below the H1 data for large positive 
$\eta(D^*)$, whereas for the ZEUS data, the NLO calculation gives a good 
description. It should be noted that the ZEUS data is compared with the recent 
ZEUS NLO QCD fit as the proton PDF in the NLO calculation. This gives a 
a somewhat larger cross section at positive $\eta(D^*)$ and somewhat 
smaller cross section at negative $\eta(D^*)$. Whether the difference in 
conclusion arises from differences in data or differences in theory is not 
clear at present. The data are, however, consistent when extrapolated to measure 
$F_2^{c\bar{c}}$ as shown in Figure~\ref{fig:f2charm}. A comparison of the cross 
sections in the same kinematic range should be made.

\begin{figure}[htp]
\begin{center}
\includegraphics[width=13.5cm]{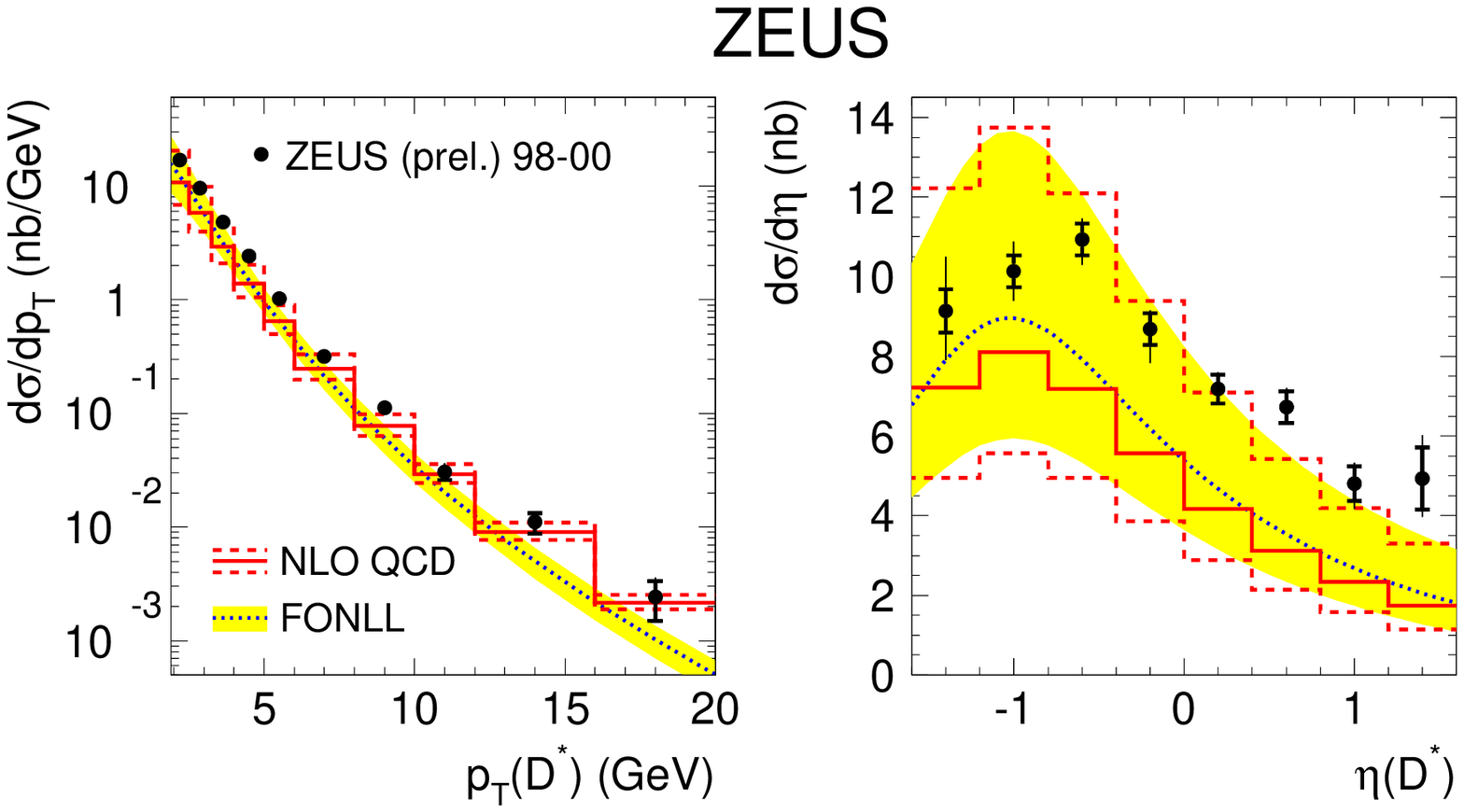}
\end{center}
\vspace{-0.5cm}
\caption{\it Charm cross sections in photoproduction at HERA compared to QCD.
\label{fig:charmx_hera} }
\end{figure}

Due to its larger cross section, charm photoproduction measurements are 
the most accurate from HERA. Theoretically, however, photoproduction has the 
additional uncertainty associated with the possibility of the photon resolving into a 
source of hadrons and the interactions behaving as a hadron-hadron 
collision (see Figure~\ref{fig:feyn1}(a)). Example data are shown in 
Figure~\ref{fig:charmx_hera} compared with NLO QCD~\cite{fmnr} and 
FONLL~\cite{fonll-hera} predictions (and NLL~\cite{nll} predictions not shown); 
none give a satisfactory representation of the data. Indeed the FONLL 
calculation which is meant to be more reliable at high $p_T(D^*)$ than 
the NLO QCD calculation gives a poorer description of the data. The precision 
of the calculations is also poor, with uncertainties as a functions of 
$\eta(D^*)$ between 30\% and 80\%, whilst the data has a precision of 
generally better than 10\%. The precision of the data will improve in time 
with more data; it is hoped that higher precision for the predictions can also 
be achieved.

\begin{figure}[htp]
\begin{center}
\includegraphics[width=13.5cm]{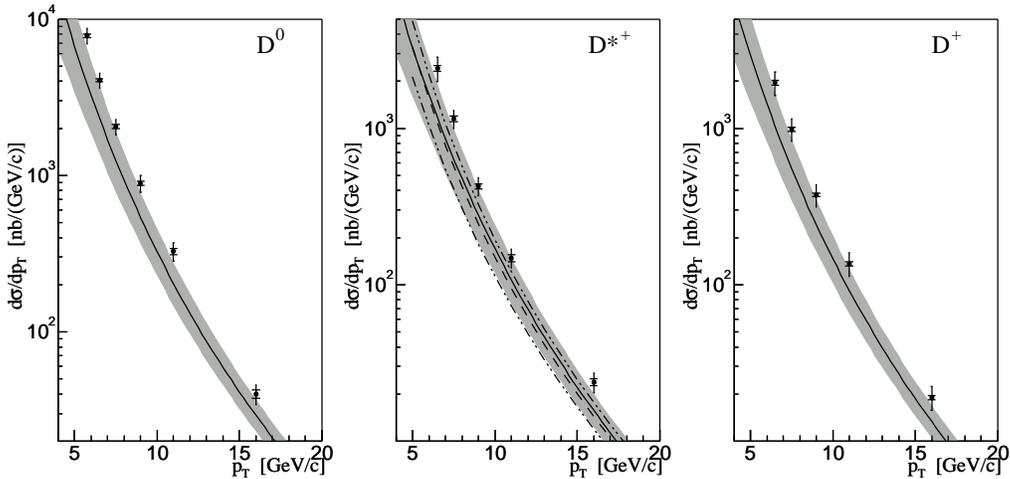}
\end{center}
\vspace{-0.5cm}
\caption{\it Measurement of charm meson cross sections from CDF compared to 
FONLL theoretical predictions.
\label{fig:charmx_tev} }
\end{figure}

Using their upgraded (vertex) detector and Run II data from the Tevatron, 
the CDF collaboration have recently made measurements of charm meson cross 
sections~\cite{charm-tev}. The data are shown in Figure~\ref{fig:charmx_tev} 
compared with FONLL and NLL calculations. The FONLL calculation employs the 
same techniques as for the calculation of $B$ meson cross sections described 
previously. The comparison between data and prediction shown in 
Figure~\ref{fig:charmx_tev} is similar to the comparison for $B$ production 
shown in Figure~\ref{fig:b_tev_1}(b). The data sample used here corresponds to 
an integrated luminosity of 6~pb$^{-1}$ which represents a very small fraction 
of what CDF hope to collect during Run II. With the increased precision and 
greatly extended kinematic range expected, these measurements will provide 
detailed comparisons with predictions and a deeper understanding of the 
dynamics of charm production.

In general, predictions of NLO QCD are below measurements of charm production, 
but compatible within the uncertainties. As the measured cross sections 
are a complicated convolution of (PDF $\otimes$ hard scatter $\otimes$ fragmentation) 
each with parameters which have associated uncertainties, e.g. scale, charm mass, 
etc., it is unclear how to improve the description of the data. Trying to 
minimise specific effects and uncertainties will help to qualify the situation. 
Examples of this are: measurements at high $p_T$ where the scale uncertainties 
are reduced; measurements of jet cross sections which are less sensitive to 
uncertainties in the fragmentation and independent measurements of the 
fragmentation in a hadron-hadron environment rather than using the 
parametrisations of LEP data.

\subsection{Measurements of beauty cross sections}

As with $D^*$ cross sections, beauty cross sections have been measured at HERA, 
LEP and the Tevatron. The measurements are, however, a complicated mix of 
different definitions both extrapolated and within the acceptance of the 
detectors.

\begin{figure}[htb]
\begin{center}
\includegraphics[width=8.5cm]{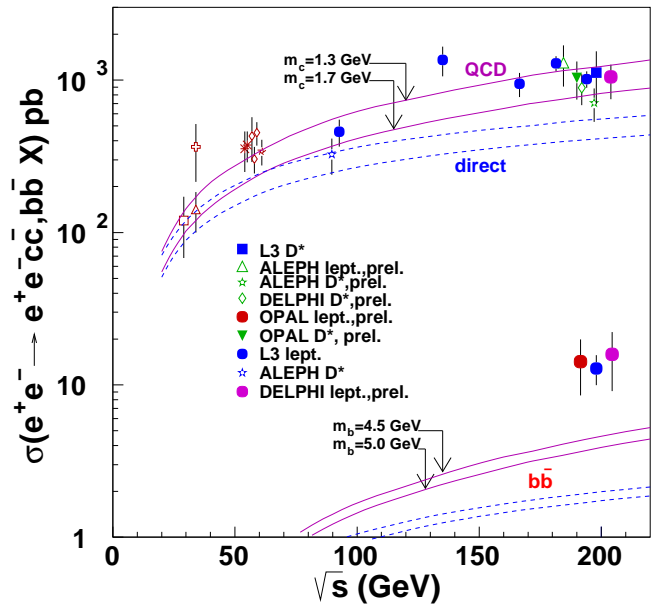}
\end{center}
\vspace{-0.5cm}
\caption{\it Charm and beauty cross sections in $\gamma \gamma$ collisions at LEP.
\label{fig:b_lep} }
\end{figure}

Figure~\ref{fig:b_lep} shows the measurements of beauty (and charm) cross sections in 
$\gamma \gamma$ collisions at LEP~\cite{lep-b}. 
All three beauty measurements lie above the 
prediction by about a factor of 3, whereas all the charm measurements are well 
described by the theory. The beauty results are similar to the first Tevatron results 
shown in Figure~\ref{fig:b_tev_1}(a). It should be noted that the LEP data have  
large extrapolation factors 
to get from the cross section measured within the acceptance of the detector to 
the total cross section shown. As said earlier, whilst providing easy comparison 
between different experiments, extrapolations to completely unmeasured angular and 
$p_T$ regions should be treated with caution. In such cases, a cross 
section in a measured kinematic region should always be given and exact details 
of the extrapolation.

\begin{figure}[htb]
\begin{center}
\includegraphics[width=6.9cm]{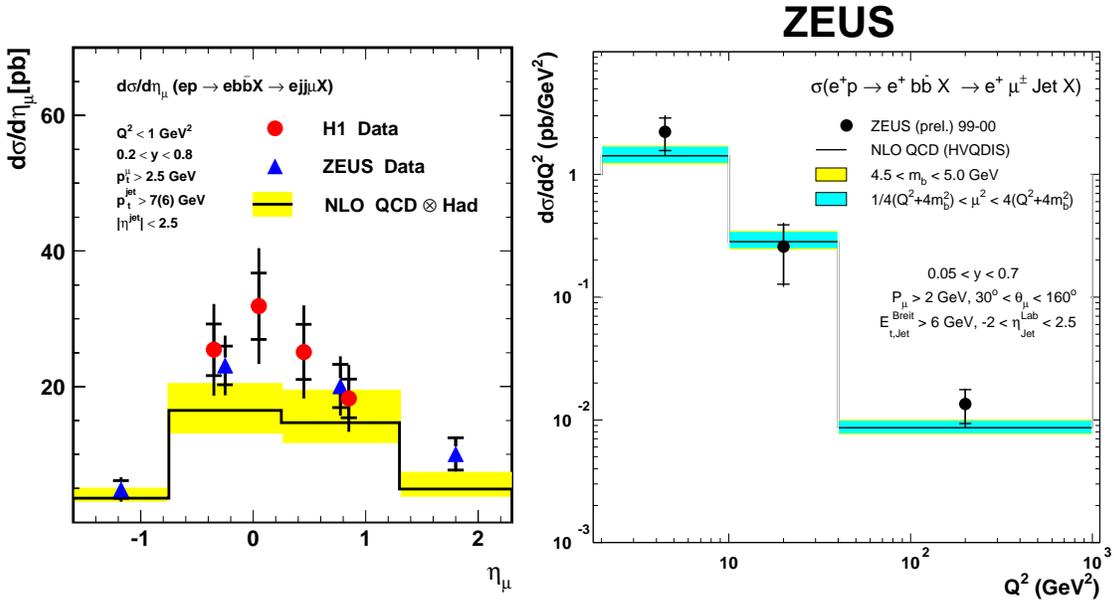}
\includegraphics[width=7.7cm]{include_figures/b_dis_q2.epsi}
\end{center}
\vspace{-0.5cm}
\caption{\it Beauty cross sections in (a) photoproduction and (b) DIS at HERA 
compared with NLO QCD.
\label{fig:b_hera} }
\end{figure}

Measurements made at HERA~\cite{hera-b} have also been a mixture of different 
styles of results. 
The latest and ``purest'' measurements made within the acceptance of the detector 
are shown compared to NLO QCD in Figure~\ref{fig:b_hera}. In 
Figure~\ref{fig:b_hera}(a) results from both experiments are shown and are 
consistent with each other. Predictions from NLO QCD are below the data, but not 
by a significant factor. This is in contrast to results which are extrapolated 
to the full phase space to all jet angles and momenta~\cite{h1b_eps03}.
A similar measurement in DIS is shown compared with NLO QCD predictions in 
Figure~\ref{fig:b_hera}(b). The prediction is again below the data, but 
consistent within the uncertainties.

Understanding in the field of beauty production is progressing 
quickly and is currently one of the most interesting challenges in collider physics. 
Since the Physics in Collision conference, the results from H1 presented in 
Figure~\ref{fig:b_hera} are new and have shown consistency between experiments and 
highlighted problems in extrapolations. Further progress from all colliders 
is to expected soon. The LEP experiments should publish their measurements. The 
HERA experiments 
should finalise the HERA I data and should receive significantly more data from 
HERA~II in the near future. The Tevatron experiments also have a wealth of data 
from Run~II to analyse. All of these future measurements and publications should 
be careful to clearly define the cross section to be measured. Extrapolations to the 
full phase-space are not intrinsically incorrect, but the initial measured cross 
section should always be quoted and the exact method of extrapolation detailed. 

\section{Conclusions}

The understanding of heavy-quark production is currently one of the most important 
challenges in QCD. In these proceedings, new results have been discussed both in 
terms of their quality and physics message. There are many technical and procedural 
issues involved in measuring heavy quarks which have to be mastered before 
the real physics can be seen. Most recent results, which provide sound measurements, 
show that although NLO QCD does a fair job in describing the data, it fails in the 
details which are now seen by the precision measurements being made. A deeper  
understanding of heavy-quark production is necessary for a more complete picture 
of QCD. It is also desirable, if not necessary, 
for future experiments such as those 
at the LHC where knowledge of the QCD background to a high precision is essential  
before physics beyond the SM can be seen. In the next few years, a combination of 
better data and improved theory should allow a detailed understanding of the 
production of heavy quarks to be achieved.

\end{document}